\documentclass{svproc}
\usepackage{url}

\usepackage{graphicx}
\usepackage{hyperref}
\usepackage{color}

\hypersetup{
    colorlinks=true,
    citecolor=blue,
    filecolor=black,
    linkcolor=black,
    urlcolor=black
}
\usepackage{footmisc}

\usepackage{color}
\usepackage[margin=1in]{geometry}
\usepackage{graphicx}
\usepackage{amsmath}
\usepackage{subcaption}

\begin{document}
\mainmatter              
\title{Design and Evaluation of a Scalable Data Pipeline for AI-Driven Air Quality Monitoring in Low-Resource Settings}
\titlerunning{Scalable Data Pipeline for AI-Driven Air Quality Monitoring}  
%
\author{Richard Sserunjogi\textsuperscript{*}\inst{1} \and Daniel Ogenrwot\textsuperscript{*}\inst{1,2} \and Nicholas Niwamanya\inst{1} \and Noah Nsimbe\inst{1} \and Martin Bbaale\inst{1} \and Benjamin Ssempala\inst{1} \and Noble Mutabazi\inst{1} \and Raja Fidel Wabinyai\inst{1} \and Deo Okure\inst{1} \and Engineer Bainomugisha\inst{1}}
\authorrunning{R. Sserunjogi, D. Ogenrwot et al.} 
%
\tocauthor{Richard Sserujongi, Daniel Ogenrwot, Nicholas Niwamanya, Noah Nsimbe, Martin Bbaale, Benjamin Ssempala, Noble Mutabazi, Raja Fidel Wabinyai, Deo Okure, and Engineer Bainomugisha}
\institute{Makerere University, Plot 56 University Pool Road, Kampala, Uganda\\
\email{richard@airqo.net, daniel@airqo.net, nicholas@airqo.net,  nsimbenoah@gmail.com, martin@airqo.net, benjamin@airqo.net, mutabazinoble@gmail.com, raja@airqo.net, dokure@airqo.net, baino@airqo.net}\\
\texttt{https://airqo.net}
\and
University of Nevada Las Vegas, Las Vegas, NV 89154, USA}

\def\thefootnote{*}\footnotetext{These authors contributed equally to this work and share first authorship.}
\def\thefootnote{\arabic{footnote}} 

\maketitle              

\begin{abstract}
The increasing adoption of low-cost environmental sensors and AI-enabled applications has accelerated the demand for scalable and resilient data infrastructures, particularly in data-scarce and resource-constrained regions. This paper presents the design, implementation, and evaluation of the AirQo data pipeline -- a modular, cloud-native Extract-Transform-Load (ETL) system engineered to support both real-time and batch processing of heterogeneous air quality data across urban deployments in Africa. It is Built using open-source technologies such as Apache Airflow, Apache Kafka, and Google BigQuery. The pipeline integrates diverse data streams from low-cost sensors, third-party weather APIs, and reference-grade monitors to enable automated calibration, forecasting, and accessible analytics. We demonstrate the pipeline’s ability to ingest, transform, and distribute millions of air quality measurements monthly from over 400 monitoring devices while achieving low latency, high throughput, and robust data availability, even under constrained power and connectivity conditions. The paper details key architectural features, including workflow orchestration, decoupled ingestion layers, machine learning-driven sensor calibration, and observability frameworks. Performance is evaluated across operational metrics such as resource utilization, ingestion throughput, calibration accuracy, and data availability, offering practical insights into building sustainable environmental data platforms. By open-sourcing the platform and documenting deployment experiences, this work contributes a reusable blueprint for similar initiatives seeking to advance environmental intelligence through data engineering in low-resource settings.
\end{abstract}

\section{Introduction}
The design and operation of scalable data pipelines are critical challenges in modern data engineering, particularly as organizations increasingly rely on heterogeneous data sources for analytics, decision-making, and the deployment of machine learning systems. Data pipelines are responsible for ingesting, transforming, and delivering structured, semi-structured, and unstructured data from diverse origins while ensuring reliability, scalability, and fault tolerance in production environment~\cite{munappy2020data,raj2020modelling,lipovac2024developing}. However, the complexity of maintaining consistent data quality and availability across varying data sources, network conditions, and operational environments presents significant engineering challenges~\cite{cai2015challenges}. Standard approaches to data ingestion and transformation often fall short when applied to resource-constrained environments, where data may arrive in varying formats and frequencies from distributed sources, including IoT devices, third-party APIs, and legacy systems operating over low bandwidth 2G GSM connectivity~\cite{bainomugisha2023design,munappy2020data}. To address these challenges, there is a growing need for modular, extensible, and resilient ETL pipelines that leverage modern workflow orchestration tools, scalable data warehousing, and decoupled architectures to support real-time and batch processing workloads.

Recent advances in Artificial Intelligence (AI), cloud-native infrastructure, and edge-based Internet of Things (IoT) systems have unlocked new capabilities for real-time monitoring and prediction in complex domains such as public health, transportation, and environmental management. In particular, air quality monitoring systems increasingly rely on AI for predictive modeling, automated sensor calibration and anomaly detection. However, real-world implementation of such intelligence demands robust and adaptive data infrastructure that can cope with the scale, velocity, and variability of environmental data, especially in low-resource settings characterized by unreliable power and intermittent internet connectivity.

This paper presents the design and evaluation of a scalable, modular, and production-grade data pipeline that underpins the \textit{AirQo Platform}, an AI-enabled air quality monitoring initiative with over 400 air quality monitors deployed across major African cities~\cite{bainomugisha2024ai,adong2022applying,bainomugisha2023design,okure2022characterization}. AirQo data pipeline aggregates measurements from a range of heterogeneous sources, including proprietary low-cost sensors~\cite{kumar2015rise}, Beta Attenuation Mass (BAM) grade reference monitors, and third-party weather APIs, to provide high-resolution environmental intelligence~\cite{sserunjogi2022seeing}. The platform serves both scientific research and public engagement objectives, supporting dashboards, APIs, and machine learning-driven forecasts~\cite{okure2025case}.

The AirQo data pipeline is built on open-source, cloud-native technologies: Apache Airflow orchestrates end-to-end workflows; Google BigQuery provides a scalable analytical warehouse; and Apache Kafka ensures real-time, decoupled streaming to downstream microservices. Through this architecture, the system seamlessly integrates both real-time and batch workflows, performs automated calibration, and feeds analytics and forecasts into public-facing and decision-support tools. The data pipeline is evaluated using both infrastructure-level and data-centric metrics: resource utilization, ingestion latency, throughput, data availability, and calibration success rate. These metrics provide insights into the trade-offs required to sustain performance under production workloads, while also meeting the reliability expectations of scientific and civic stakeholders. Our findings demonstrate how principled data engineering, grounded in modular design and open technologies, can drive sustainable innovation even under stringent constraints.

\noindent This paper makes the following contributions

\begin{itemize}
    \item We present the design and implementation of a modular, scalable ETL pipeline capable of integrating heterogeneous data sources and supporting AI-driven workflows.
    \item We present a detailed performance evaluation of the pipeline in production based on resource metrics, data quality, and operational reliability.
    \item We articulate practical engineering lessons on observability, resilience, and architectural trade-offs for practitioners building real-world data infrastructure in constrained environments.
    \item We provide the complete source code and documentation of the AirQo data pipeline and platform as an open-source resource to enable collaboration and extension of the pipeline for similar environmental monitoring applications.
\end{itemize}

\section{Related Work}\label{sec:label}
In this section, we review prior research on scalable data pipelines, intelligent environmental data systems, and the unique challenges of deploying data infrastructure in low-resource settings. These related works provide critical context for our study, which lies at the intersection of modern data engineering and environmental sensing.

\subsection*{Scalable Data Pipeline Architectures}
Modern data pipelines are designed to handle increasing volumes of diverse data in real-time and batch modes. The increasing demand for timely and actionable insights from large-scale data has led to the development of numerous frameworks and architectures for scalable data pipelines. Workflow orchestration platforms such as Apache Airflow~\cite{apacheairflow,harenslak2021data}, Luigi~\cite{luigi}, and Prefect~\cite{prefect} have been employed to automate ETL workflows, enabling organizations to manage complex data dependencies while maintaining observability and fault tolerance. These tools provide the backbone for building modular and maintainable pipelines, supporting both batch and streaming data processing use cases. 

Cloud-based data warehousing solutions like Google BigQuery~\cite{bigquery}, Amazon Redshift~\cite{redshift}, and Snowflake~\cite{snowflake} have transformed the landscape of data storage and analytics, providing scalable and cost-efficient infrastructures capable of handling massive datasets with low latency. These platforms are essential components in modern data engineering architectures, allowing engineers to perform interactive analytics and support machine learning workflows at scale.

Prior studies have explored scalable pipelines for IoT data~\cite{iotpipelines}, focusing on data ingestion and processing using cloud services for high-throughput scenarios. However, many existing approaches do not document the practical integration of heterogeneous sensor types, operational considerations in resource-constrained environments, and the end-to-end architecture required for consistent, reliable data delivery to downstream analytics and decision-making systems.

\subsection*{Intelligent Environmental Data Integration and Systems}
Recent efforts have highlighted the need for intelligent data integration in health and environmental applications. For instance, Ndembi et al.~\cite{ndembi2025integrating} proposed a data lakehouse architecture for integrating environmental and health data in African cities. Similarly, Chhikara et al.~\cite{chhikara2021federated} explored federated learning pipelines for environmental sensing but acknowledged the challenge of robust data availability and quality in under-connected regions. In another domain, Ullah et al.~\cite{ullah2024ai} implemented a low-cost, AI-enhanced air quality monitoring system using ESP32 microcontrollers, highlighting hardware-level innovations but with limited focus on end-to-end data workflows.

In the domain of environmental data processing, initiatives like OpenAQ~\cite{openaq} and AirVisual~\cite{airvisual} aggregate air quality data from diverse sources to support research and public policy. These projects demonstrate the utility of centralized repositories for environmental data, but often rely on periodic batch updates and may lack advanced ETL orchestration for real-time applications. Similarly, IoT-based environmental monitoring systems~\cite{iotenv} have leveraged low-cost sensors and cloud services to enable real-time environmental monitoring, yet challenges remain in ensuring data quality, consistency, and the reliable handling of diverse sensor data streams in production environments. 

Machine learning techniques are increasingly employed to calibrate low-cost air quality sensors by learning from co-located reference monitors and incorporating auxiliary features like temperature and humidity~\cite{adong2022applying,hashmy2023modular,ali2024leveraging,ali2024machine}. These models help mitigate sensor drift and improve data reliability. AirQo adopts a similar approach by integrating a prediction and calibration microservice within its pipeline that operates on both streaming and historical datasets. This allows for near real-time correction of raw sensor data, supporting both public dashboards and research analytics.

\subsection*{Data Infrastructure Challenges in Low-Resource Settings}

Building robust data pipelines in low-resource settings entails navigating several constraints including unreliable internet connectivity, limited computational and energy resources, and infrastructural constraints. Recent research highlights how intermittent network conditions can lead to frequent data gaps, complicating time-series continuity and reducing the reliability of real-time applications~\cite{pinder2019opportunities,ahmed2018internet,bainomugisha2023design}. Techniques such as opportunistic data uploading, edge-based caching, and delayed backfilling pipelines have been proposed to address this challenge~\cite{zhalgasbekova2017opportunistic,bainomugisha2023design}. The AirQo pipeline adopts these approaches by embedding automated historical backfilling DAGs and leveraging fault-tolerant message queues (Kafka) to decouple data producers and consumers. 

Several air quality monitoring platforms in the Global South have turned to solar-powered sensors and energy-efficient data collection firmware to maintain consistent operation under fluctuating voltage or extended outages~\cite{bainomugisha2023design,zafra2024designing,dushyanth2025design,mberu2025urban}. Moreover, constrained cloud access has driven interest in hybrid edge–cloud architectures, wherein localized edge nodes perform pre-processing and buffering before periodic upstream transmission~\cite{higashino2017edge,bainomugisha2023design}. Low-resource settings also suffer from limited availability of high-quality ground truth data, which complicates model training and sensor calibration. This has led to innovations in transfer learning, synthetic data generation, and federated calibration models~\cite{chhikara2021federated,endres2022synthetic}. Such techniques attempt to reduce dependence on site-specific labeled datasets while preserving generalizability. 

Taken together, these contributions outline a growing research frontier that seeks to democratize environmental sensing infrastructure in resource-constrained environments. Our work builds on these efforts by presenting the design, implementation, and operational evaluation of a scalable ETL pipeline capable of ingesting and transforming heterogeneous data sources in near real-time using open-source and cloud-based technologies. Using the AirQo Data Pipeline as a case study, we demonstrate practical strategies for building robust data pipelines to support environmental monitoring and analytics in low-resource settings, providing actionable insights for practitioners aiming to deploy scalable data infrastructure in similar domains.

\section{AirQo Data Pipeline}
\begin{figure}[ht]
    \centering
    \includegraphics[width=\textwidth]{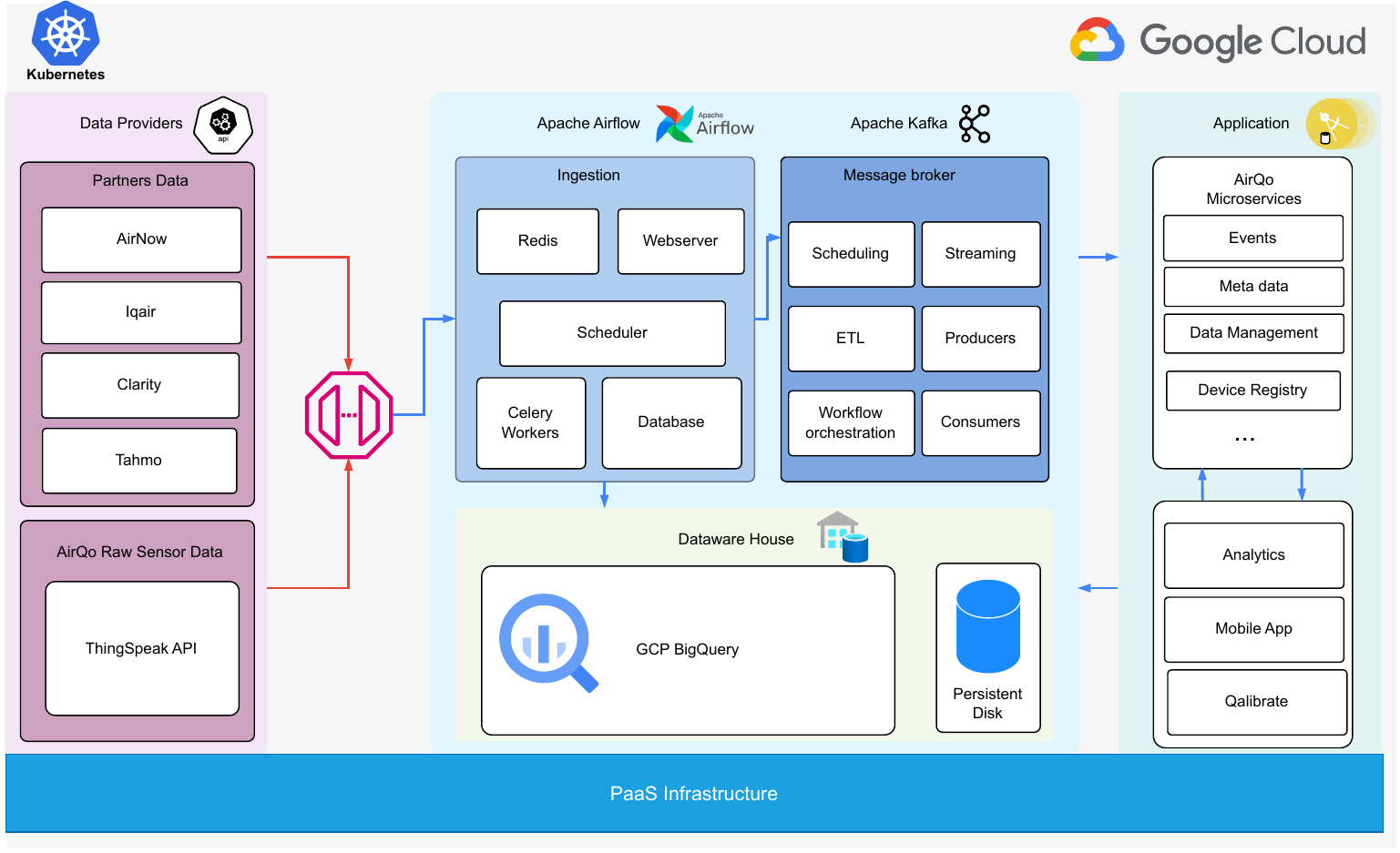}
    \caption{AirQo data pipeline architecture overview.}
    \label{fig:pipeline_architecture}
\end{figure}

The AirQo data pipeline serves as a practical application of modern data engineering principles to environmental monitoring in low-resource urban environments. Designed for modularity, scalability, and reliability, the pipeline integrates heterogeneous data sources including low-cost sensors, BAM reference-grade monitors, and weather data APIs into a cohesive architecture capable of supporting near real-time and batch analytics. The pipeline is designed as a cloud-native, modular system designed for scalability, fault tolerance, and real-time analytics. As illustrated in Figure~\ref{fig:pipeline_architecture}, the pipeline integrates diverse data sources, orchestrates ingestion and processing workflows using Apache Airflow, manages asynchronous communication with Apache Kafka, and provides scalable storage and analytics through Google Cloud Platform (GCP) services.

\subsection{Data Processing Workflows}
\begin{figure}[ht]
    \centering
    \includegraphics[width=\textwidth]{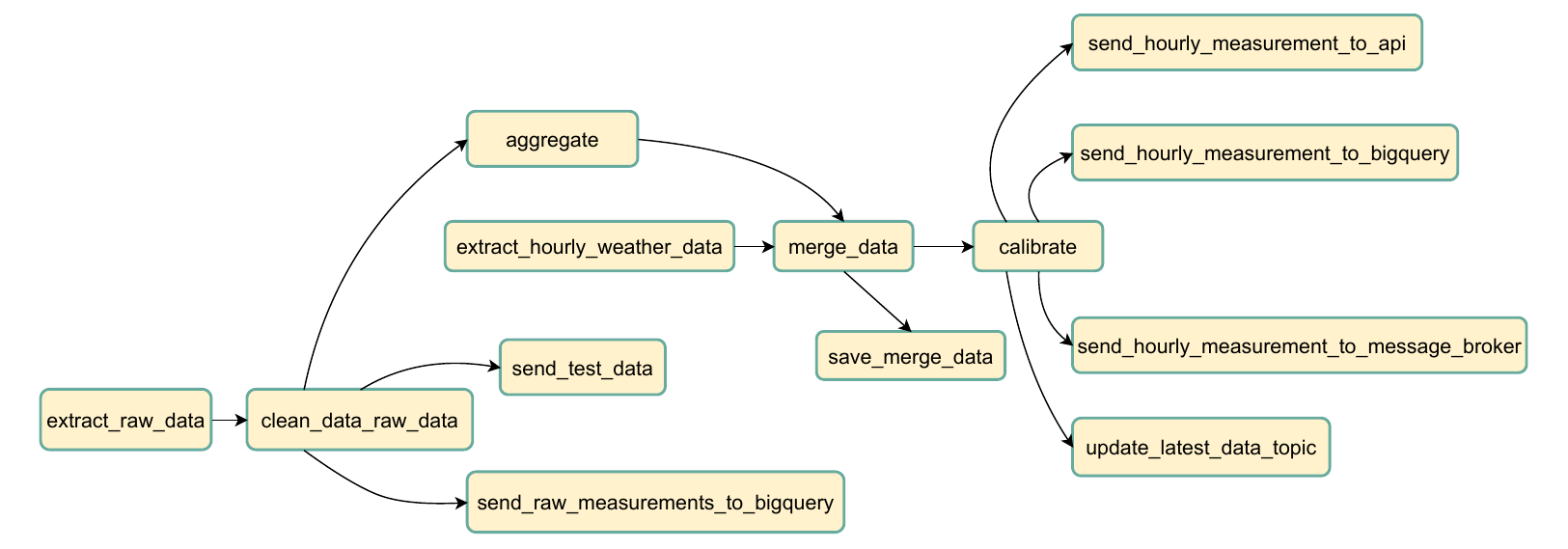}
    \caption{Directed Acyclic Graph (DAG) for AirQo near real-time low-cost measurement workflow. Each node represents a distinct ETL task, including raw data extraction, cleaning, calibration, aggregation, and multi-channel distribution to BigQuery, Apache Kafka, and external APIs. Edges indicate task dependencies and execution order.}
    \label{fig:airqo_dag}
\end{figure}

\noindent \textbf{Data Sources.} The AirQo platform ingests data from a diverse mix of internal sensors and external collaborators. Internally, it relies on a network of proprietary low-cost sensors and Beta Attenuation Monitors (BAMs) to generate high-frequency air quality readings. To enhance spatial coverage and improve data reliability, this core network is supplemented with data from third-party providers including IQAir, MetOne, and Clarity. Additionally, meteorological data is sourced through public weather APIs including TAHMO~\cite{van2014trans} and OpenWeatherMap ~\cite{openweathermap}. The ingestion layer supports a multi-tenant design that allows seamless integration of external data streams, enabling continuous synchronization with partner APIs. AirQo’s own devices communicate via the ThingSpeak API~\cite{thingspeak}, which acts as an aggregation endpoint for field-deployed sensors. These heterogeneous data streams, which vary in format, precision, and update frequency form the basis for downstream analytics, forecasting, and operational monitoring. Once captured, all raw input is forwarded to the next stage in the pipeline for transformation and validation.
\\~\\
\noindent \textbf{Data Ingestion.} The AirQo pipeline uses Apache Airflow~\cite{Singh2019,haines2022workflow} as its primary orchestration tool for managing data ingestion and transformation tasks. Airflow enables the scheduling, execution, and monitoring of ETL workflows through modular scripts written in Python. Each workflow is implemented as a Directed Acyclic Graph (DAG), which defines a series of interdependent tasks responsible for extracting data from various sources, performing necessary cleaning and validation, and loading the transformed outputs into AirQo's data storage systems. Airflow's architecture consists of several core components: a scheduler that manages task execution timing, Celery workers for parallel and distributed task execution, a metadata database to track workflow states, and a webserver providing a user interface for visualization and monitoring. The use of DAGs ensures deterministic and fault-tolerant execution, allowing the system to reliably manage data ingestion across heterogeneous sources and formats. This orchestration framework enables AirQo to operate a mix of real-time, hourly, and historical ingestion workflows efficiently, each optimized for the specific timing and transformation requirements of its corresponding data stream. Figure~\ref{fig:airqo_dag} illustrates the DAG used in orchestrating the real-time low-cost air quality measurement workflow within the AirQo data pipeline. The DAG defines a sequence of interdependent tasks, beginning with the extraction and cleaning of raw sensor data, followed by hourly aggregation and calibration using synchronized weather data. Subsequent tasks are responsible for dispatching the processed measurements to various destinations, including BigQuery for archival and analytics, Apache Kafka for streaming-based microservice consumption, and the AirQo API for real-time user access. The DAG structure ensures fault-tolerant, ordered task execution and facilitates parallelization where dependencies allow, enabling consistent performance under dynamic data loads. Table~\ref{tab:realtime_lc_measurements} summarizes the task, functionality and dependencies of this DAG. For brevity, we have not included several other DAGs defined in the AirQo data pipeline.

\begin{table}[ht]
  \renewcommand{\arraystretch}{1.5} 
\centering
\caption{AirQo near real-time low-cost measurement DAG tasks}
\begin{tabular}{clp{6.8cm}c}
\hline
\textbf{\#} & \textbf{Task Name} & \textbf{Functionality} & \textbf{Dependencies} \\
\hline
1 & \texttt{extract\_raw\_data} & Extracts raw data for the last hour from low-cost devices. & None \\
2 & \texttt{clean\_data\_raw\_data} & Cleans and structures the raw data by removing outliers and enforcing data type checks. & 1 \\
3 & \texttt{save\_test\_data} & Saves the cleaned data for testing. & 2 \\
4 & \texttt{aggregate} & Resamples device data to an hourly frequency. & 2 \\
5 & \texttt{extract\_hourly\_weather\_data} & Extracts hourly weather data. & None \\
6 & \texttt{merge\_data} & Merges hourly air quality data with hourly weather data. & 4, 5 \\
7 & \texttt{calibrate} & Calibrates the merged data. & 6 \\
8 & \texttt{send\_hourly\_measurements\_to\_api} & Sends calibrated data to AirQo API. & 7 \\
9 & \texttt{send\_hourly\_measurements\_to\_message\_broker} & Sends calibrated data to the MessageBroker. & 7 \\
10 & \texttt{send\_hourly\_measurements\_to\_BigQuery} & Sends calibrated data to BigQuery. & 7 \\
11 & \texttt{update\_latest\_data\_topic} & Updates the latest data topic in the MessageBroker. & 7 \\
\hline
\end{tabular}
\label{tab:realtime_lc_measurements}
\end{table}

\noindent \textbf{Message Broker.} To decouple data ingestion from downstream processing, the AirQo data pipeline leverages Apache Kafka~\cite{garg2013apache,wang2015building}, a distributed, high-throughput, publish-subscribe messaging system. Apache Kafka serves as the central communication backbone between Apache Airflow and various microservices, enabling scalable, real-time streaming and asynchronous event-driven architectures. Once data has been transformed and validated, it is published to specific Kafka topics. These topics act as temporary data stores, allowing microservices to consume messages at their own pace and according to their individual service-level requirements. This approach enhances pipeline resilience by insulating data producers from consumer availability or latency. Apache Kafka's architecture allows for persistent buffering, ensuring that transient failures in downstream systems do not result in data loss. To reduce payload size and transmission latency, messages are serialized and transmitted in compact byte formats. Additionally, Apache Kafka topics are protected using authentication mechanisms, with access credentials securely managed via Google Secret Manager. This separation of secrets from the runtime environment reinforces security across the ingestion and distribution layers of the pipeline.
\\~\\
\noindent \textbf{Data Warehousing.} The AirQo data pipeline leverages Google BigQuery as its centralized, serverless data warehousing solution, designed to support scalable analytical queries over large volumes of air quality and meteorological data. As a fully managed platform, BigQuery enables low-latency querying, schema evolution, and robust integration with machine learning pipelines and dashboarding tools. Processed and calibrated data from Apache Airflow workflows is stored in BigQuery in well-structured datasets, supporting both batch analytics and real-time insights. The core datasets include (i) \textit{raw data}, containing minimally processed sensor readings; (ii) \textit{averaged data}, with hourly and daily aggregates for trend analysis; (iii) \textit{consolidated data}, which merges device measurements, weather readings, and site metadata into unified records; and (iv) \textit{forecast data}, used for predictive modeling. This schema facilitates spatial-temporal queries across diverse sensor networks, including low-cost devices and reference-grade monitors. Access to the warehouse is tightly governed using Google Cloud’s role-based access control (RBAC), service accounts, and metadata tagging. Downstream applications include anomaly detection, air quality forecasting, public dashboards, and regulatory compliance reporting, all powered by curated datasets from BigQuery.

\subsection{Microservices and Data Distribution}

The AirQo platform employs a microservices-based architecture to enable scalable, modular access to processed air quality and meteorological data. These stateless services consume data from Apache Kafka topics and the BigQuery warehouse to serve multiple downstream applications, ensuring flexibility and performance across both real-time and historical use cases. Each microservice is designed with a single responsibility principle, which improves maintainability and facilitates independent deployment. Key services include:

\begin{itemize}
    \item \textbf{Device Registry Service:} Maintains up-to-date metadata on sensors, including device ID, geolocation, firmware version, and operational status.
    
    \item \textbf{Event Service:} Captures internal platform events, such as data availability, processing anomalies, and API access logs, supporting observability and auditability.
    
    \item \textbf{Metadata Service:} Provides contextual data including site names, sensor elevations, city identifiers, and partner associations, which is critical for both visualization and analytics.
    
    \item \textbf{Data Management Service:} Consumes calibrated and aggregated air quality data from Kafka topics and delivers it to various endpoints, including mobile apps, public dashboards, and other AirQo APIs.
    
    \item \textbf{Calibrate Service:} A specialized machine learning-enabled microservice that supports the calibration of raw measurements using reference-grade monitors and weather features. It interfaces with both streaming (Kafka) and batch (BigQuery) data layers.
    \item \textbf{Predict Service:} A forecasting microservice that leverages trained machine learning models to generate short-term predictions of air quality metrics, enhancing real-time decision-making and enabling proactive environmental response.
\end{itemize}

\noindent These microservices operate in a decoupled manner, communicating through RESTful APIs and Kafka topics. Real-time consumers subscribe to Kafka topics to ingest high-frequency data updates, while analytical services perform SQL-based queries on BigQuery for aggregated insights. To ensure secure data distribution, all services are deployed with authenticated access tokens, enforced via role-based access control (RBAC) in Kubernetes. Message payloads transmitted through Kafka are serialized (e.g., using Avro or JSON) to minimize network overhead. Load balancing and horizontal scaling are implemented using Kubernetes’ native mechanisms to handle variable traffic from external clients and internal processing jobs. This architecture supports AirQo’s commitment to real-time environmental intelligence by providing reliable, scalable, and secure access to high-quality data across various digital platforms.

\subsection{Deployment and Operational Considerations}

The AirQo data pipeline is deployed using a containerized, cloud-native architecture on Google Cloud Platform (GCP), leveraging Kubernetes for workload orchestration and scaling. The infrastructure is organized into two isolated Kubernetes clusters: a production cluster responsible for live operations and a staging cluster designated for pre-deployment integration testing. This separation supports robust version control, enables staged rollouts, and reduces the risk of system-wide failures. Each cluster runs on 4 VM nodes configured with 4 vCPUs, 16 GB of memory, and 200 GB of persistent disk space per node. One node is dedicated as controller and the rest configured as worker nodes. Cluster networking is managed using Calico as the Container Network Interface (CNI), chosen for its high performance, network policy enforcement, and flexibility compared to alternatives such as Flannel. Public-facing microservices are exposed via an NGINX ingress controller, which is routed through an HAProxy load balancer to manage HTTP(S) traffic securely and efficiently. Operational management follows a GitOps paradigm, using Argo CD to maintain declarative configuration and automate continuous delivery. Prometheus and Grafana are integrated into the pipeline to provide real-time metrics, system health dashboards, and custom alerting rules. CI/CD pipelines are defined using GitHub Actions, ensuring automated testing, container builds, and environment-specific deployments with Helm charts. Logs from Apache Airflow, Apache Kafka, and microservices are aggregated and routed to centralized logging backends for observability and fault tracing. This deployment strategy ensures high availability, modular scalability, and operational resilience, which are essential for sustaining uninterrupted ingestion, processing, and distribution of large-scale environmental datasets within the AirQo ecosystem.

\section{Data Quality and Analytics}

In this section, we evaluate the data quality and discuss analytics feature of the AirQo pipeline. Ensuring the quality and integrity of the collected data is critical to achieving the AirQo’s mission of delivering actionable environmental insights. The system tracks multiple data quality metrics across all stages of the pipeline to monitor availability, uniqueness, and calibration success rate.

\subsection{Quality Metrics}

 \subsubsection{Data Availability:} 
\begin{figure}
    \centering
    \includegraphics[width=0.9\linewidth]{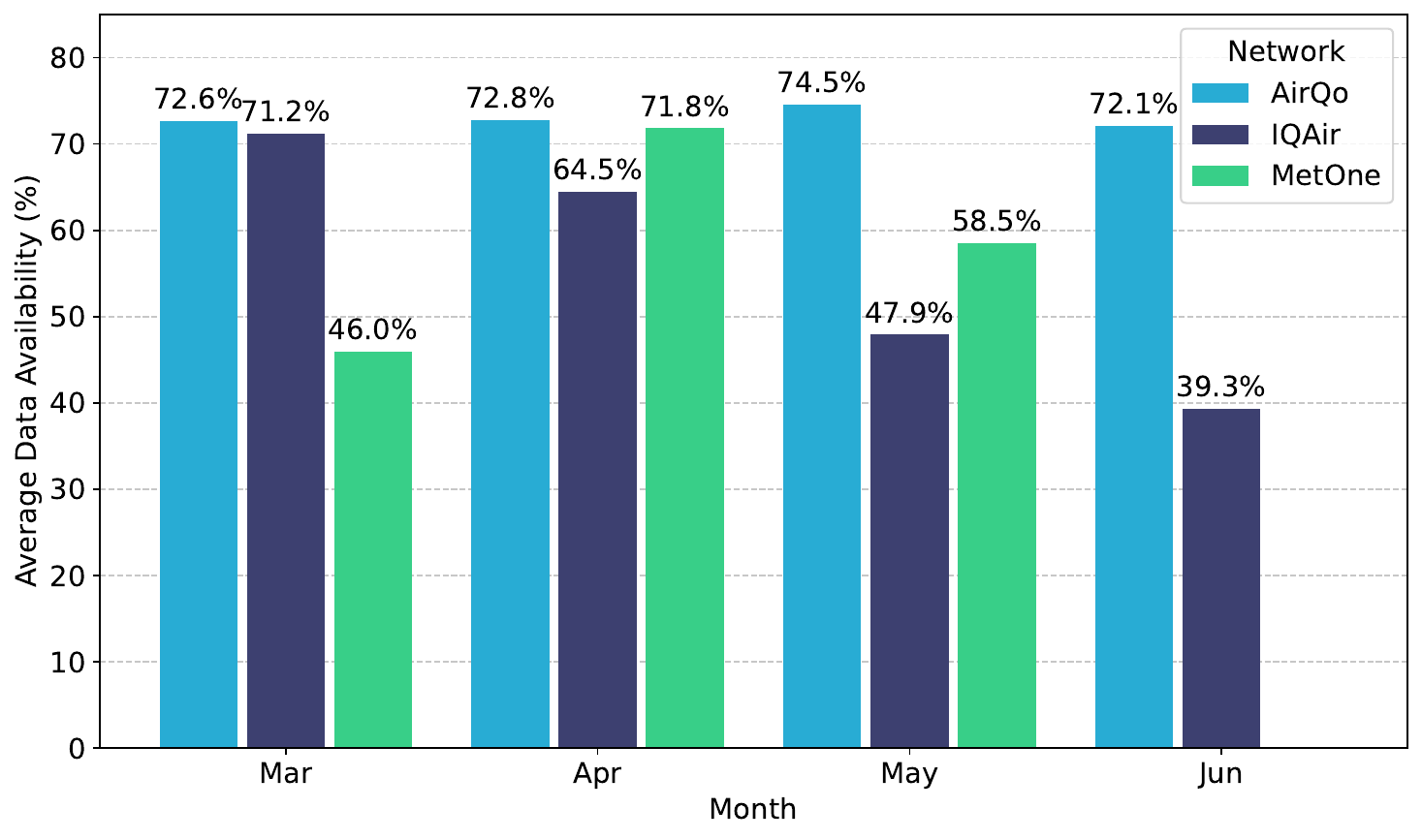}
    \caption{Monthly average data availability (in percentage) for AirQo, IQAir, and MetOne sensor networks from March to June 2025. Data availability reflects the proportion of hours in which devices reported valid measurements, serving as a key indicator of sensor uptime and transmission reliability.}
    \label{fig:data-availability-trends}
\end{figure}

 Measures the proportion of expected hours in which a device reports data. This helps identify data gaps caused by hardware failures, power outages, or connectivity issues.
 \begin{equation*}
     \text{Availability Rate} = \left(\frac{\text{Hours with Data}}{\text{Total Hours}}\right) \times 100
 \end{equation*}
\noindent An analysis of average data availability from March to June 2025, based on data ingested into the AirQo pipeline, reflected distinct but complementary patterns for the three air quality sensor data vendors i.e., AirQo, IQAir, and MetOne as shown in Figure~\ref{fig:data-availability-trends}. During this period, AirQo devices consistently maintained high availability, averaging above 70\%, with a slight decline from 72.81\% in April to 72.08\% in June. This stability reflects the robustness of the AirQo pipeline and its capacity for real-time ingestion and timely recovery via backfilling mechanisms. In contrast, data from the IQAir devices supported in the AirQo pipeline showed a sharp decline in availability from 64.5\% in April to 47.91\% in May, and further down to 39.3\% in June. This downward trend has been reported to the sensor network operator, i.e., the Permian Health Lung Institute (PHLI) team for further investigation.
MetOne devices, which are reference-grade monitors, experienced a more moderate decline, dropping from 71.81\% in April to 58.47\% in May. This is because some of the monitors were decommissioned following administrative directives. These metrics demonstrate AirQo’s data pipeline comparative strength in maintaining consistent data streams despite operating in low-resource environments. The system’s architecture characterized by scheduled retries, modular ingestion, and Kafka-based buffering contributes to this resilience. Such insights inform ongoing efforts to optimize deployment strategies, improve fault detection, and ensure reliable air quality monitoring across all networks.

\subsubsection{Calibration Rate:} 
\begin{figure}[ht]
    \centering
    \begin{subfigure}[b]{0.49\textwidth}
        \centering
        \includegraphics[width=\linewidth]{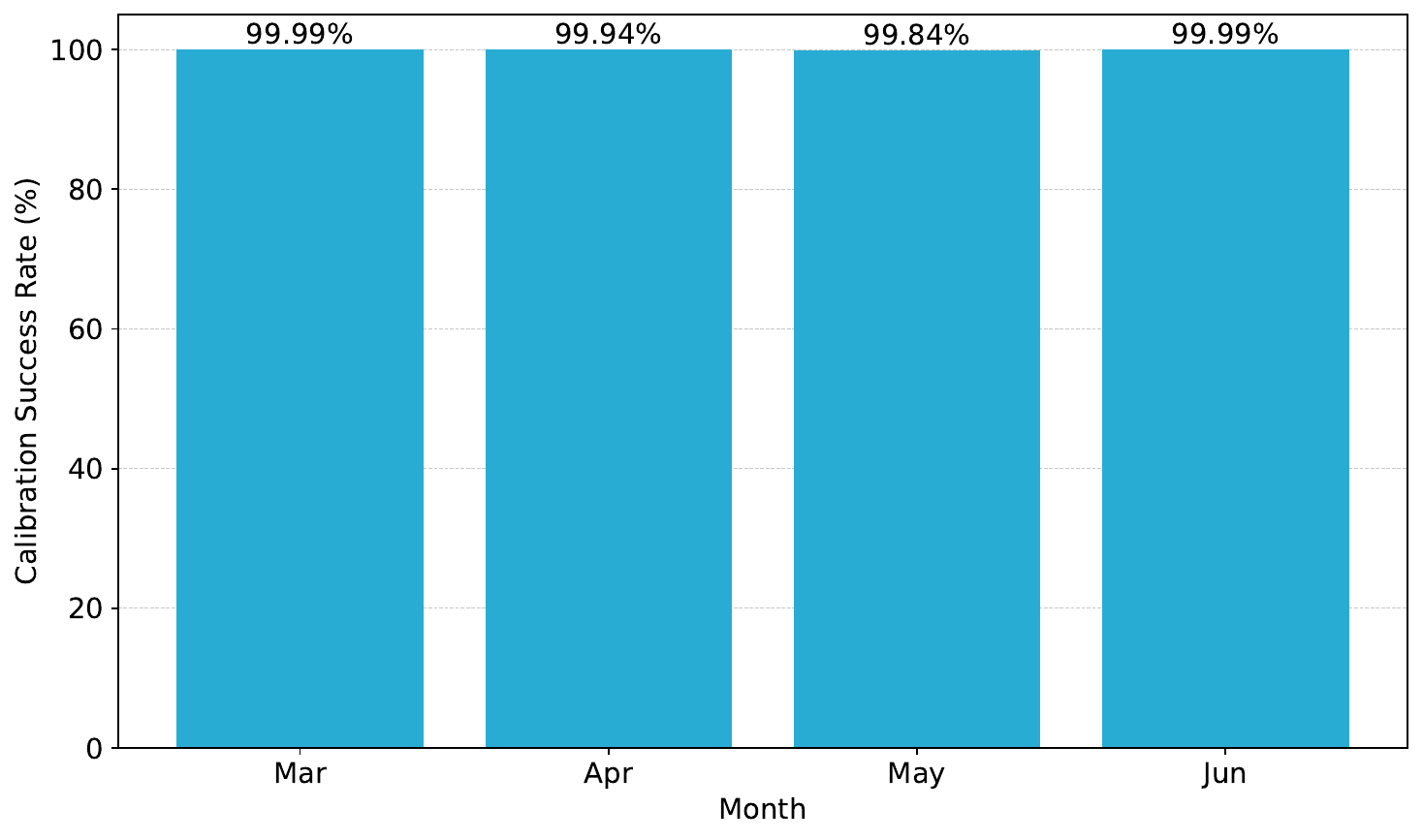}
        \caption{Monthly calibration success rate across the network.}
        \label{fig:calibration-success}
    \end{subfigure}
    \hfill
    \begin{subfigure}[b]{0.49\textwidth}
        \centering
        \includegraphics[width=\linewidth]{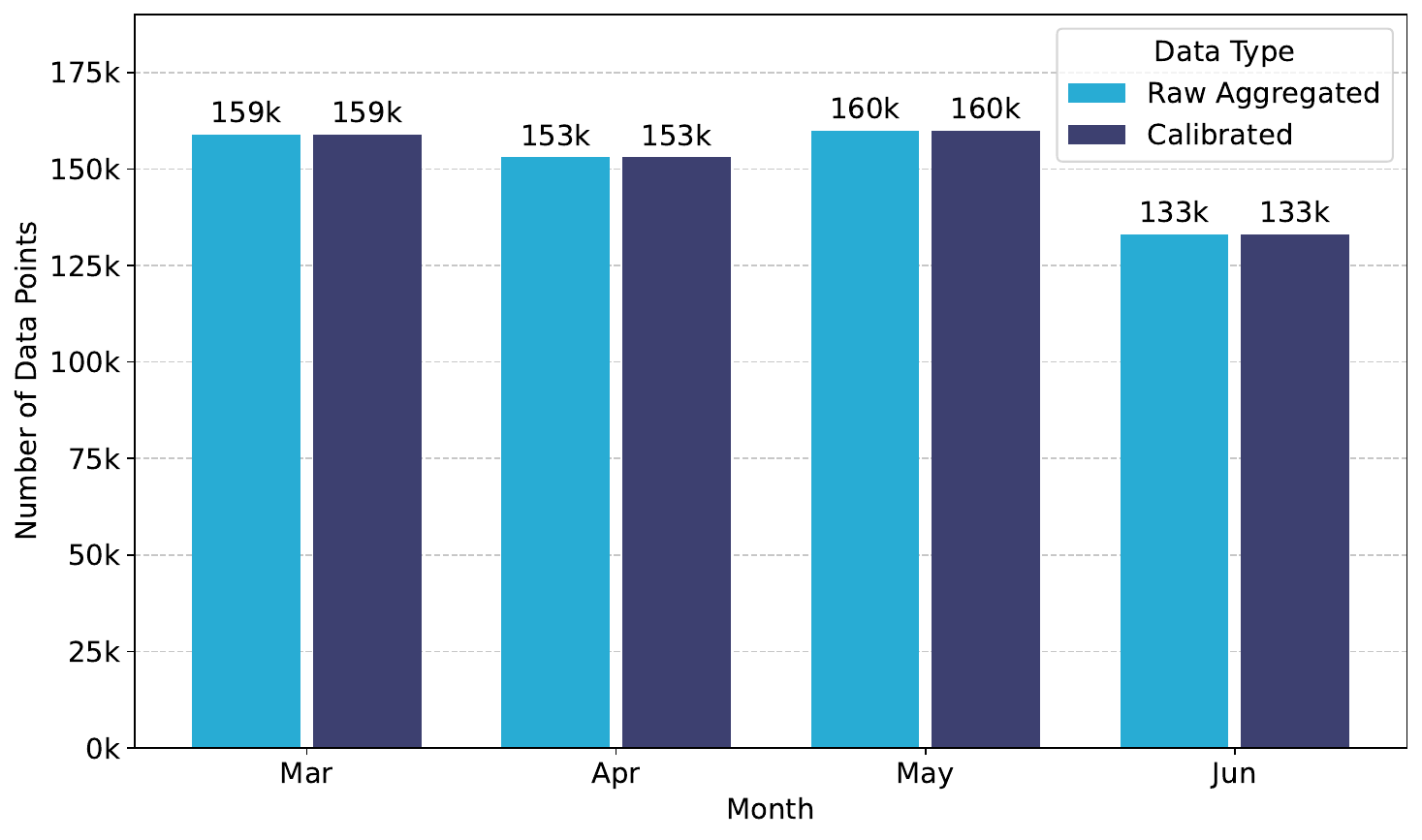}
        \caption{Monthly comparison of raw vs. calibrated data points.}
        \label{fig:raw-vs-calibrated}
    \end{subfigure}
    \caption{Evaluation of calibration pipeline performance from March to June 2025. Left: success rates of calibration. Right: parity between raw and calibrated hourly data volume, highlighting throughput and consistency.}
    \label{fig:calibration-performance}
\end{figure}

Assesses the ratio of raw data that has been successfully calibrated. This metric helps evaluate the pipeline's ability to apply calibration models consistently. 
\begin{equation*}
     \text{Calibration Rate} = \left(\frac{\text{Hours with Calibrated Data}}{\text{Hours with Raw Data}}\right) \times 100
\end{equation*}
\noindent During the second quarter of 2025, the AirQo data pipeline maintained near-perfect calibration success rates, highlighting the reliability of its data correction workflows as shown in Figure~\ref{fig:calibration-success}. From March through June, the average calibration rate consistently exceeded \textbf{99.9\%}, with exact monthly values of 99.99\% (March), 99.94\% (April), 99.84\% (May), and 99.99\% (June). These represent the proportion of raw sensor measurements that were successfully processed and transformed into calibrated outputs using model-based corrections within the ETL pipeline. This performance reflects the robustness of the calibration module embedded in the DAG workflows managed by Apache Airflow, which includes validation, anomaly filtering, and model inference steps. The consistently high calibration rate suggests minimal pipeline disruptions, accurate execution of calibration models, and effective fallback mechanisms for handling edge cases. Furthermore, this reinforces the system’s capacity to deliver reliable, actionable environmental data particularly in urban deployments where low-cost sensors may be subject to drift, cross-sensitivity, or fluctuating environmental conditions. The success of the calibration process directly enhances the quality of air quality analytics and forecasting, ensuring data consistency across time and space. This reliability is vital for downstream applications including public dashboards, mobile apps, and policy advisory tools.

To further assess calibration effectiveness, we examined the volume of raw versus calibrated hourly data points during the period from March to June 2025 as shown in Figure~\ref{fig:raw-vs-calibrated}. The plot indicates a near 1:1 correspondence between raw aggregated and calibrated data points, with each month recording over 150,000 measurements in both categories. Specifically, March and May reached the highest volumes at 159k and 160k respectively, while April maintained parity with 153k data points in both raw and calibrated forms. June saw a slight decline to 133k, yet still maintained full calibration coverage for the available raw data. This alignment highlights the efficiency and scalability of the AirQo calibration pipeline, which successfully processed almost all raw sensor readings without backlog or significant drop-off. The pipeline’s ability to maintain such parity at scale demonstrates the reliability of the calibration models and the robustness of the data transformation workflows orchestrated by Apache Airflow.

\subsection{Analytics}

To deliver actionable insights to diverse stakeholders including researchers, policymakers, environmental agencies, and field technicians, the AirQo platform incorporates a dual-layer analytics system. This system facilitates interactive exploration, decision support, and predictive modeling based on curated environmental datasets. First, a self-service analytics interface is provided through Apache Superset, enabling users to explore high-value metrics such as device uptime, calibration trends, and data availability among others. Superset dashboards empower stakeholders to rapidly detect anomalies, monitor long-term trends, and identify gaps in data collection without requiring direct access to raw data or underlying infrastructure.  Second, the platform includes a custom-built analytics service~\footnote{\url{https://analytics.airqo.net}} designed to extend beyond visualization. This component supports advanced analytical tasks such as geospatial sensor placement recommendations, historical forecasting of pollutant levels, trend analysis, network management, collocation configuration and targeted diagnostics for device behavior. By integrating machine learning workflows directly into the analytics layer, AirQo enables dynamic and adaptive decision-making processes that are tightly coupled with the evolving realities of urban air quality management in low-resource contexts.

\section{Infrastructure Performance}
Evaluating the operational performance of the AirQo data pipeline is critical to demonstrating its scalability and reliability for processing heterogeneous data streams in production environments.
\subsection{Resource Management and Scalability}
To assess resource consumption and system stability, we analyzed monitoring logs from the AirQo production Kubernetes cluster where the data pipeline is deployed over a continuous 30-day period, from June 1, 2025 to June 30, 2025. This dataset provided visibility into CPU, memory, and network traffic utilization.

\begin{figure}[ht]
    \centering
    \begin{minipage}{0.49\textwidth}
        \centering
        \includegraphics[width=\linewidth]{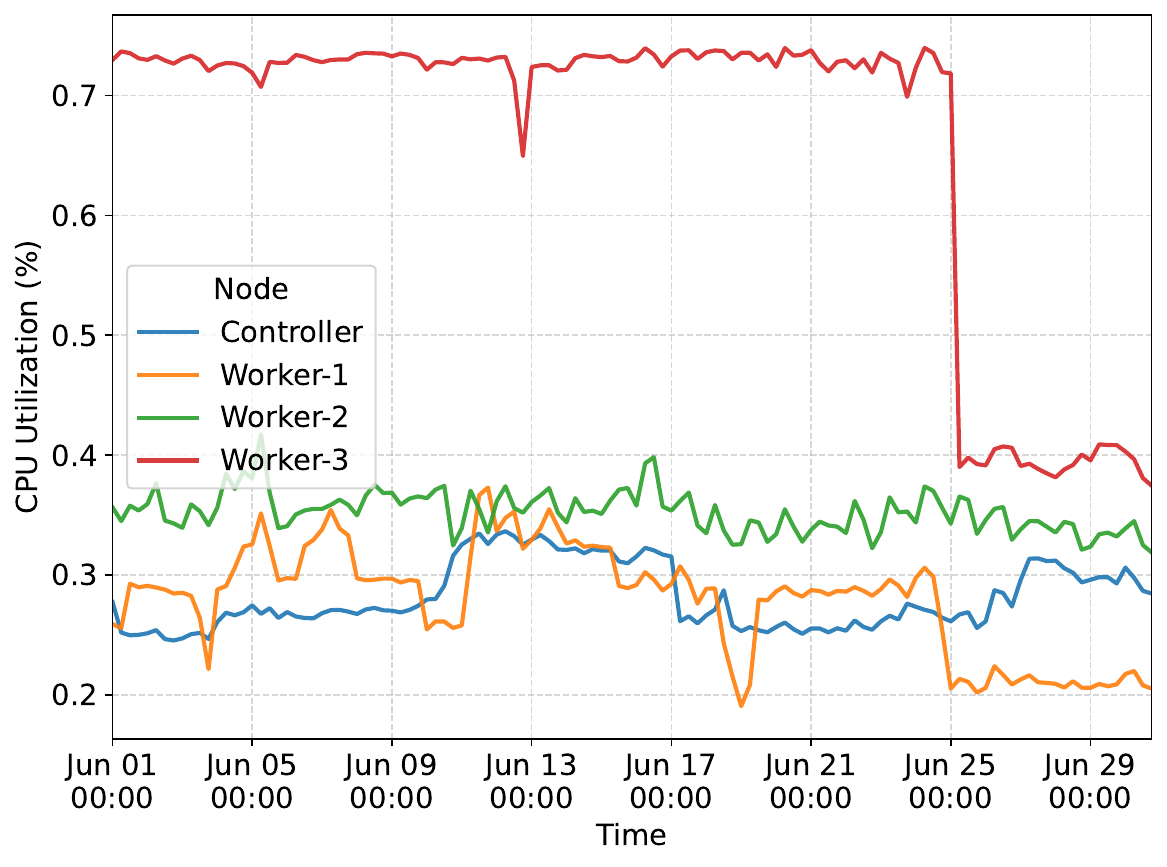}
        \caption{CPU utilization trends across AirQo cluster nodes from the 1\textsuperscript{st}-30\textsuperscript{th} June, 2025.}
        \label{fig:cpu_utilization}
    \end{minipage}
    \hfill
    \begin{minipage}{0.49\textwidth}
        \centering
        \includegraphics[width=\linewidth]{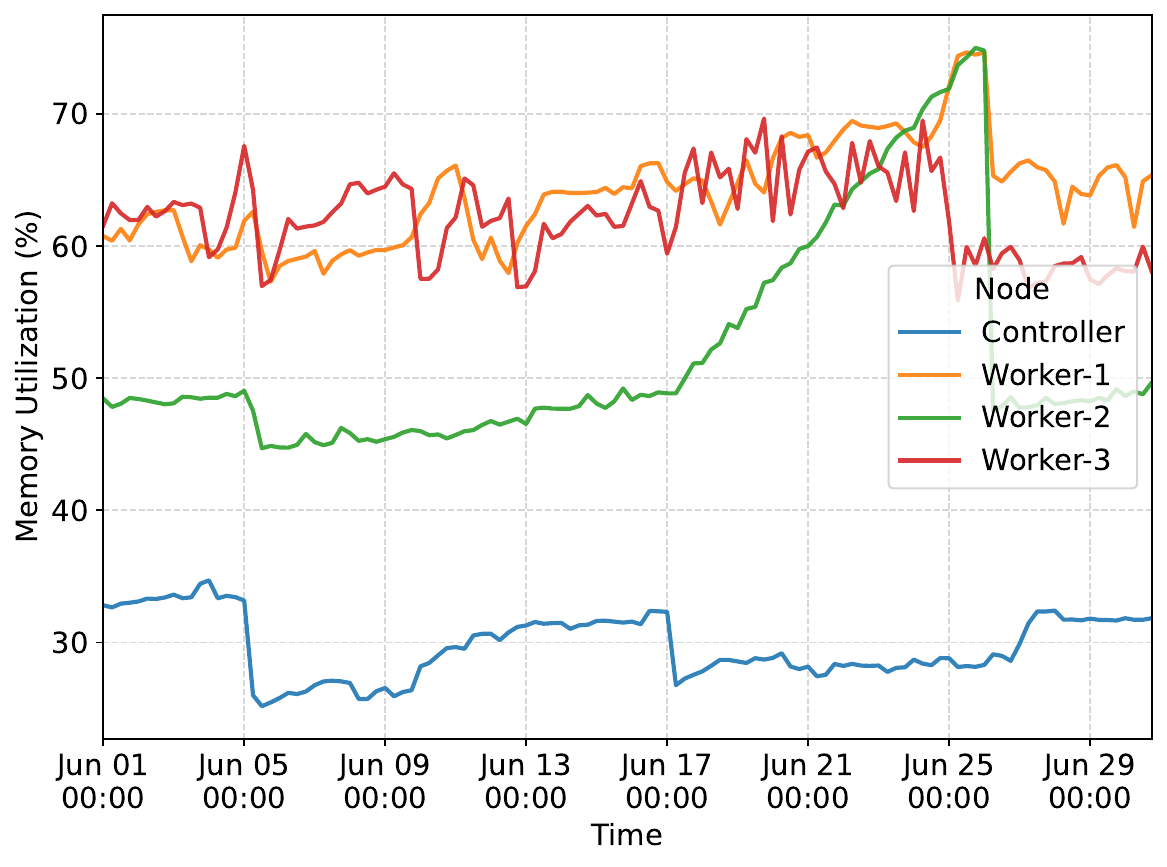}
        \caption{Memory utilization trends across AirQo cluster nodes from the 1\textsuperscript{st}-30\textsuperscript{th} June, 2025.}
        \label{fig:memory_utilization}
    \end{minipage}
\end{figure}

\textbf{CPU Utilization.}
Figure~\ref{fig:cpu_utilization} illustrates the comparative CPU utilization across the AirQo cluster nodes from  1\textsuperscript{st} to 30\textsuperscript{th} June, 2025. The Controller node shows consistently low and stable CPU activity, consistent with its coordination role. In contrast, Worker-3 experiences significantly higher and more volatile CPU usage, frequently exceeding 70\% and occasionally nearing 100\%. This pattern suggests that Worker-3 is either overburdened by processing tasks or handling a disproportionate share of computational load. Worker-1 and Worker-2 demonstrate more balanced and moderate usage, indicating effective task distribution. These findings highlight a potential bottleneck and suggest that load balancing or scaling measures may be needed for Worker-3 to ensure system reliability.

\textbf{Memory Utilization.}
Figure~\ref{fig:memory_utilization} presents memory utilization over time for each node. The Controller maintains a relatively low and consistent memory footprint, reinforcing its role as a lightweight orchestrator. Among the workers, Worker-3 again exhibits the highest memory consumption, with consistently elevated levels suggesting a heavier or more memory-intensive workload. This sustained high memory usage may impact system responsiveness or lead to out-of-memory risks under increased demand. Worker-1 and Worker-2 show similar utilization curves, suggesting equitable resource usage. Overall, the results indicate that Worker-3 may require resource scaling or workload optimization to maintain balanced performance across the cluster.

\begin{figure}[ht]
    \centering
    \begin{subfigure}[b]{0.49\linewidth}
        \includegraphics[width=\linewidth]{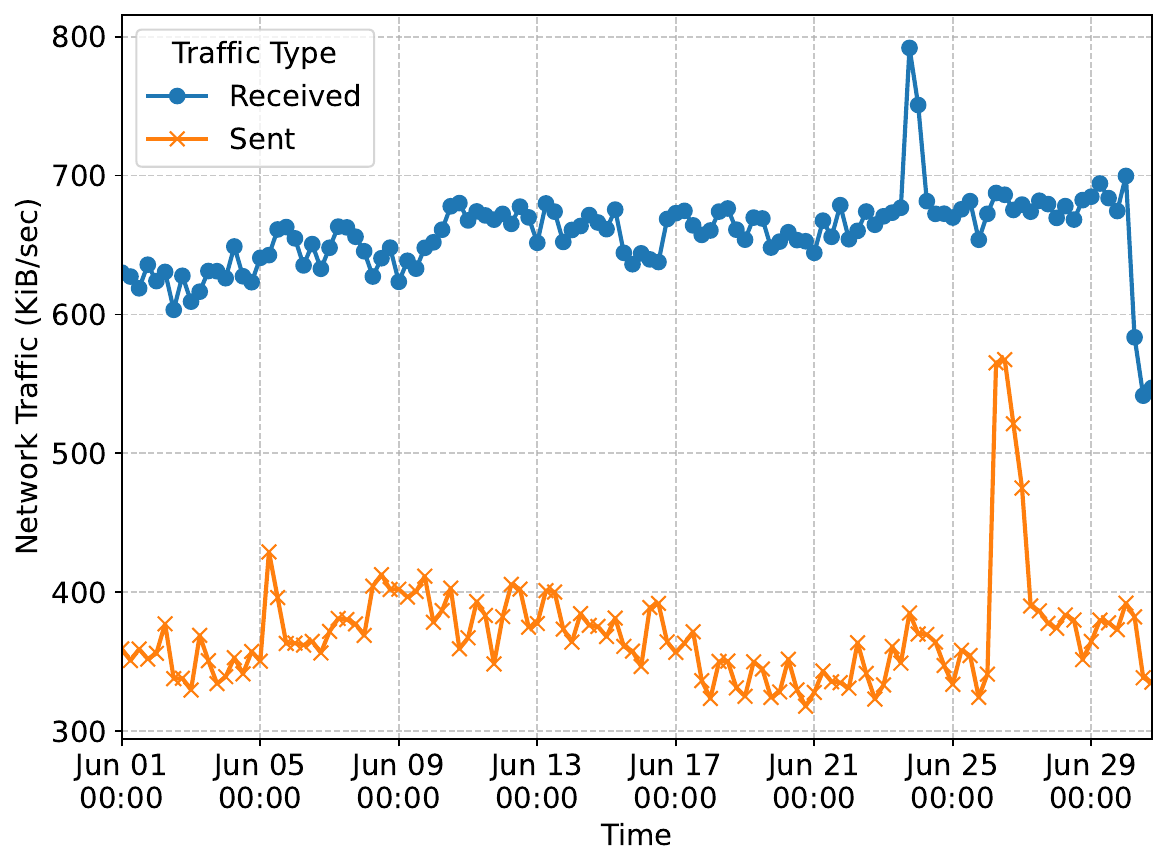}
        \caption{Controller.}
        \label{fig:controller}
    \end{subfigure}
    \hfill
    \begin{subfigure}[b]{0.49\linewidth}
        \includegraphics[width=\linewidth]{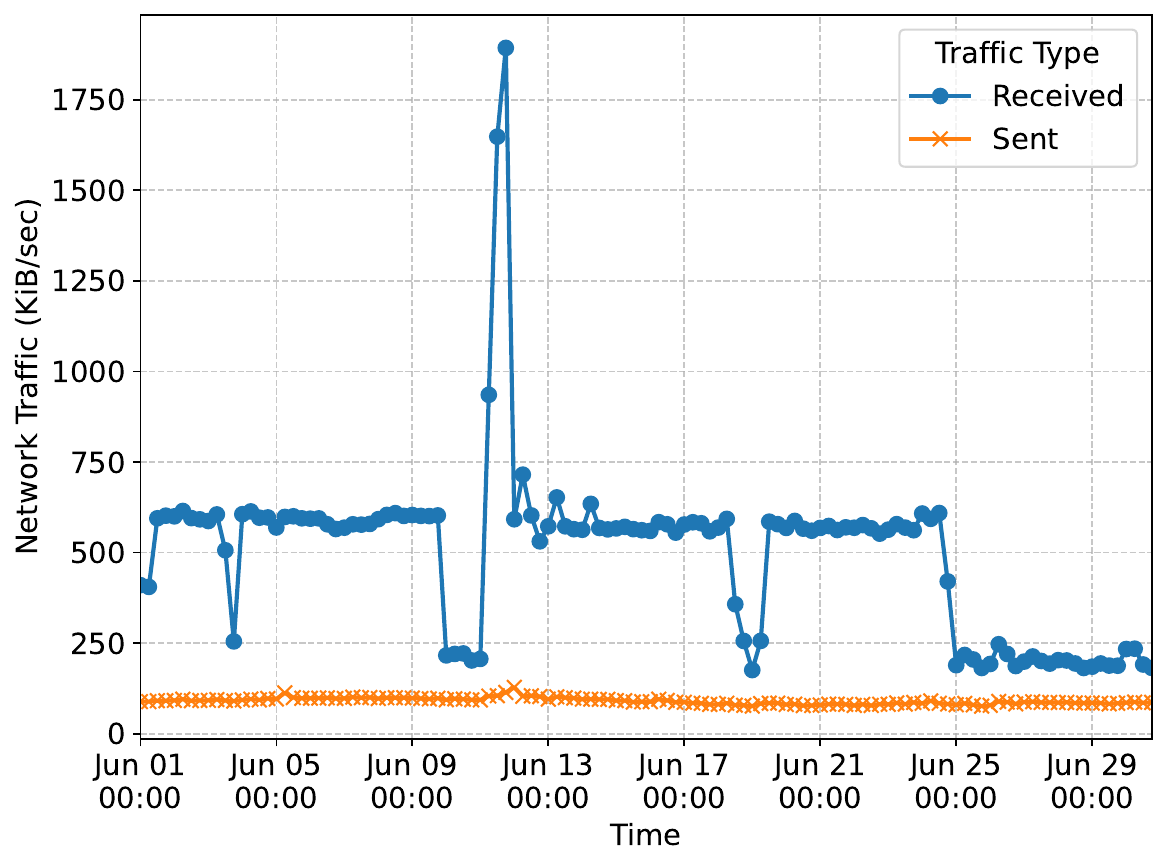}
        \caption{Worker-1.}
        \label{fig:worker-1}
    \end{subfigure}
    \hfill
    \begin{subfigure}[b]{0.49\linewidth}
        \includegraphics[width=\linewidth]{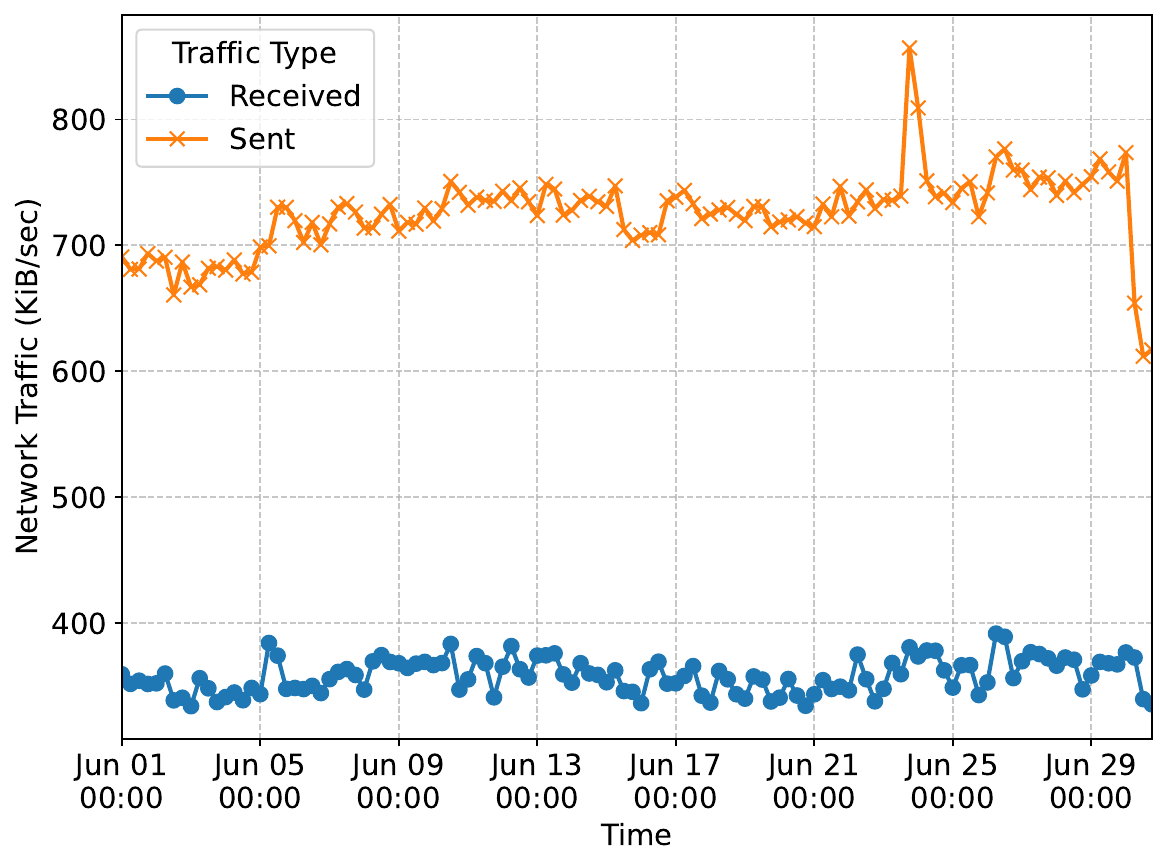}
        \caption{Worker-2.}
        \label{fig:worker-2}
    \end{subfigure}
    \hfill
    \begin{subfigure}[b]{0.49\linewidth}
        \includegraphics[width=\linewidth]{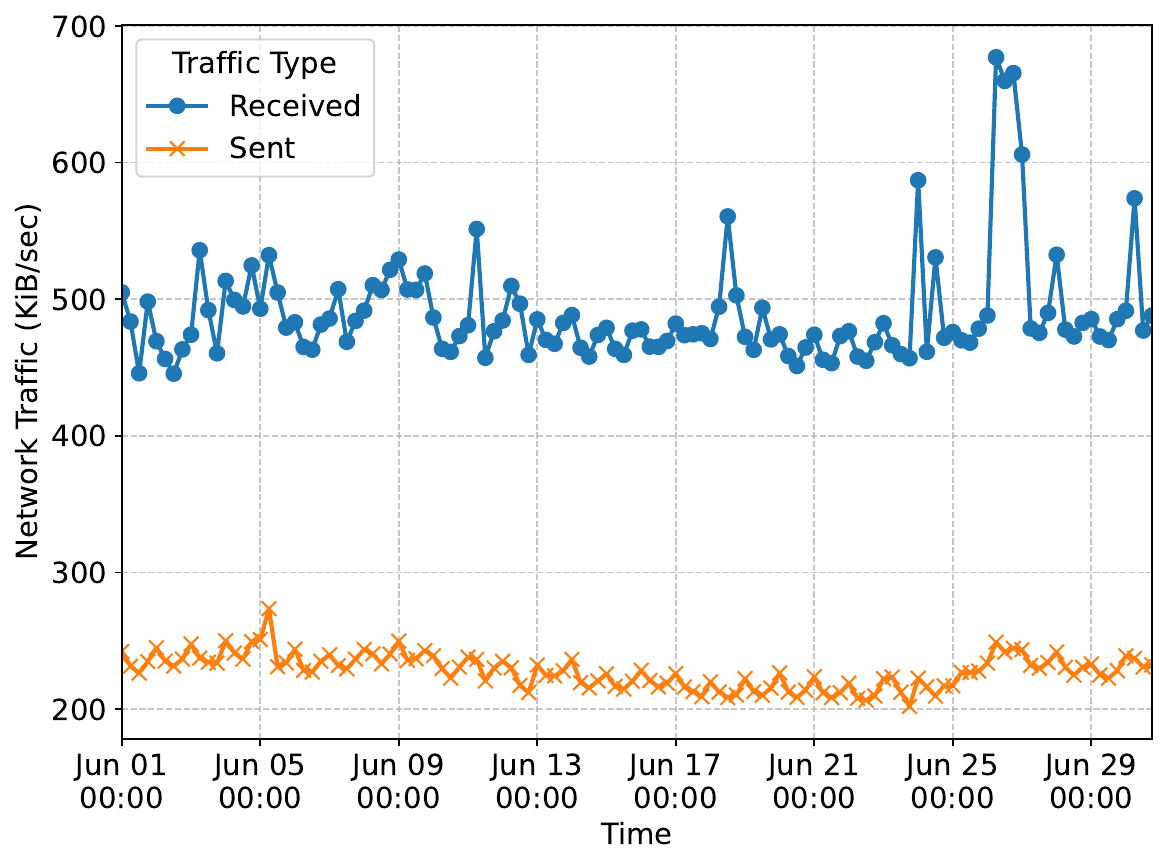}
        \caption{Worker-3.}
        \label{fig:worker-3}
    \end{subfigure}
    \caption{Network traffic pattern across AirQo production cluster nodes for the time period 1\textsuperscript{st} - 30\textsuperscript{th} June, 2025. Each figure shows the received and sent traffic in KiB/sec, highlighting the communication dynamics of the controller and worker nodes.}
    \label{fig:network_traffic_all_nodes}
\end{figure}

\textbf{Network Traffic Patterns.} Figure~\ref{fig:network_traffic_all_nodes} presents a comparative analysis of network traffic across the production cluster nodes. The \texttt{Controller} node exhibits a consistent pattern of high incoming traffic (averaging ~650 KiB/s) and relatively lower outgoing communication, consistent with its orchestration role and periodic synchronization with sensors and workers. Worker-1 shows the most volatile pattern, with spikes in received traffic exceeding 1800 KiB/s and intermittent dropouts, suggesting bursty upstream data ingestion or potential network congestion. Worker-2 demonstrates consistently high outgoing traffic, often exceeding 750 KiB/s, suggesting it may be responsible for relaying aggregated sensor data or distributing inference results downstream. In contrast, Worker-3 shows consistently high incoming traffic averaging at 500 KiB/s with random spikes not exceeding 700 KiB/s. These insights reveal traffic asymmetry across the cluster and suggest that dynamic load balancing or node-specific optimization may improve bandwidth utilization and resilience across the AirQo data pipeline.

\subsection{Throughput and Data Volumes}
The AirQo data pipeline ingests and processes approximately \textbf{~5.76 million} raw air quality data entries per month. This estimate is based on an average of 20 data points per hour per device, collected across roughly 400 active sensors, operating continuously over 720 hours in a typical month. These data points span multiple pollutant measurements (e.g., PM$_{2.5}$, PM$_{10}$), environmental parameters (e.g., temperature, humidity), and device metadata, forming a rich multivariate time-series dataset. Ingestion rates typically peak during batch device data uploads and scheduled synchronizations with partner APIs. To sustain high-throughput ingestion while avoiding bottlenecks, the system uses Apache Airflow with a \texttt{CeleryExecutor}, which enables distributed, concurrent execution of DAG tasks. 

\noindent Overall, the performance evaluation demonstrates that the AirQo data pipeline meets the requirements for scalable, reliable, and low-latency ingestion and processing of heterogeneous environmental data streams. The architecture supports high data availability and throughput while remaining adaptable to evolving workload requirements, making it a practical reference design for similar environmental monitoring and data engineering applications.

\section{Lessons Learned}

The design, deployment, and operation of the AirQo data pipeline offer valuable lessons for the broader data engineering and environmental sensing communities, particularly those operating in resource-constrained or infrastructure-fragmented settings. These insights are grounded in real-world system behavior and serve to inform both architectural choices and operational strategies for similar initiatives. Our experience reinforces that combining modular open-source technologies with cloud-native design principles enables robust, scalable, and maintainable data infrastructures in emerging contexts.

\begin{enumerate}
   \item \textbf{Modular Design Enables Flexibility.} Structuring the pipeline as a collection of modular Airflow DAGs, each representing a distinct stage such as ingestion, transformation, or calibration, enabled rapid iteration and smooth integration of new data sources. This design allowed individual components to be developed, tested, and maintained independently, reducing the risk of system-wide failures and improving team productivity. The ability to isolate and evolve specific parts of the pipeline proved essential for scalability and long-term maintainability.

    \item \textbf{Decoupling Enhances Resilience.} Integrating Apache Kafka as an asynchronous communication layer between upstream and downstream components proved pivotal. Kafka’s publish-subscribe architecture ensured that ingestion continued uninterrupted during consumer-side delays, decoupling throughput constraints and allowing independent scaling of producers and consumers. This design not only increased fault tolerance but also enabled robust microservice orchestration at scale.

   \item \textbf{Observability is Foundational.} Building comprehensive observability into the pipeline through structured logging, Apache Airflow DAG monitoring, Slack-based alerting, and real-time metric dashboards proved critical for maintaining operational reliability. These tools significantly reduced mean time to recovery (MTTR), enabling the team to shift from reactive issue resolution to proactive system health monitoring. This not only improved service uptime but also ensured data trustworthiness.

    \item \textbf{Backfilling Mechanisms are Essential.} In low-resource environments, interruptions due to sensor downtime, power outages, or external API failures are common and often unpredictable. Incorporating automated historical backfilling into the daily pipeline operations proved essential for maintaining continuous data records. This approach eliminated the need for manual recovery efforts, preserved the integrity of long-term datasets, and ensured that the system could reliably support both scientific research and data-driven policy interventions.

    \item \textbf{Leverage Cloud Scalability.} Leveraging Google BigQuery for analytical workloads and Kubernetes for deployment orchestration offered operational elasticity with minimal overhead. The system scaled naturally with increasing sensor footprints and partner integrations, without significant re-engineering. This elasticity also enabled parallel experimentation with forecasting models and anomaly detection workflows.

    \item \textbf{Trade-offs Between Real-Time and Batch Processing.} While operational dashboards and real-time alerting demanded low-latency data pipelines, tasks like calibration and hourly aggregation benefited from batched execution. Recognizing this trade-off allowed us to tailor compute strategies per workflow, optimizing both cost and performance without compromising data quality or availability.

    \item \textbf{Caching Improves Efficiency and Reduces Latency.} Introducing lightweight caching layers for frequently accessed metadata and calibration models significantly reduced redundant queries to upstream services and database lookups. By leveraging in-memory caches (e.g., Redis) within key microservices, the pipeline achieved lower response times and reduced compute loads on BigQuery and external APIs. 
\end{enumerate}

\noindent These lessons represent the practical insights gained from deploying and maintaining a city-scale environmental data system over an extended period. They reflect not only the technical viability of building scalable and dependable pipelines in low-resource settings, but also the importance of thoughtful architectural choices and strong operational practices to ensure long-term success.

\section{Conclusion}
This paper has presented the design and evaluation of a scalable, modular, and production-grade data pipeline that powers the AirQo Platform, an AI-driven air quality monitoring system deployed across major African cities. Leveraging open-source, cloud-native technologies including Apache Airflow, Kafka, and Google BigQuery, the system supports both real-time and batch processing of heterogeneous environmental data from low-cost sensors, reference monitors, and third-party APIs. Our evaluation demonstrates that the pipeline achieves a data availability rate of approximately 70\% and a calibration success rate exceeding 99.9\% over a three-month period, even under sensor operational constraints of unreliable internet connectivity, and intermittent power. These performance outcomes emphasize the effectiveness of principled data engineering practices in sustaining reliable environmental intelligence pipelines in low-resource settings. The deployment experience has yielded several key lessons: the importance of modular design for flexible evolution, Kafka-based decoupling for system resilience, structured observability for rapid incident response, and automated backfilling to mitigate data gaps. In addition, strategic use of caching, cloud elasticity, and workflow-specific processing models enabled operational robustness. Future work includes integrating generative-AI capabilities into the analytics layer, deploying edge-based preprocessing, and adopting adaptive pipeline scheduling to further enhance autonomy and reduce cloud dependencies. By open-sourcing the platform and contributing empirical insights, we aim to inform and accelerate similar efforts to deploy scalable environmental data systems in emerging regions.

\subsubsection{Data Availability.}
The datasets analyzed in this study can be made available upon reasonable request to the authors. To promote transparency and reproducibility, the complete source code for the AirQo data pipeline including data ingestion, transformation, distribution, and orchestration is publicly accessible on GitHub under an open-source license\cite{airqo-platform}. This repository also includes deployment scripts and documentation for the associated infrastructure, enabling researchers and practitioners to replicate or extend the pipeline for similar environmental monitoring applications.

\subsubsection{Funding.}
This work was supported by Google.org grant 1904-57882, EPSRC/GCRF grant EP/T00343X/1, Belgium through the Wehubit programme implemented by Enabel Wehubit Grant Agreement BEL1707111-AP-05-2, and U.S Mission grant \# SUG50021CA3041. The opinions, findings and conclusions stated herein are those of the authors and do not necessarily reflect those of the funders.

\subsubsection{\ackname}
The authors would like to appreciate the feedback and input to the research paper from the AirQo team members, collaborators, and partners of the AirQo research project including researchers, community members, and government stakeholders.


%
%
\bibliographystyle{splncs04}
\bibliography{references}

\begin{thebibliography}{10}
\providecommand{\url}[1]{\texttt{#1}}
\providecommand{\urlprefix}{URL }
\providecommand{\doi}[1]{https://doi.org/#1}

\bibitem{adong2022applying}
Adong, P., Bainomugisha, E., Okure, D., Sserunjogi, R.: Applying machine learning for large scale field calibration of low-cost pm2. 5 and pm10 air pollution sensors. Applied AI Letters  \textbf{3}(3), ~e76 (2022)

\bibitem{ahmed2018internet}
Ahmed, N., De, D., Hussain, I.: Internet of things (iot) for smart precision agriculture and farming in rural areas. IEEE internet of things journal  \textbf{5}(6),  4890--4899 (2018)

\bibitem{apacheairflow}
AirFlow, A.: Apache airflow. Available at \url{https://airflow.apache.org/} (Accessed on August 19, 2025) (2025)

\bibitem{airqo-platform}
AirQo: Airqo platform github source code. Available at \url{https://github.com/airqo-platform/} (Accessed on August 19, 2025) (2019)

\bibitem{airvisual}
AirVisual: Airvisual: Real-time air quality monitoring. Available at \url{https://www.iqair.com/air-quality-monitors} (Accessed on August 19, 2025) (2025)

\bibitem{ali2024machine}
Ali, M.S.: Machine learning based calibration techniques for low-cost air quality sensors: thesis for doctor of philosophy, electronic and computer engineering, massey university  (2024)

\bibitem{ali2024leveraging}
Ali, S., Alam, F., Potgieter, J., Arif, K.M.: Leveraging temporal information to improve machine learning-based calibration techniques for low-cost air quality sensors. Sensors  \textbf{24}(9), ~2930 (2024)

\bibitem{iotenv}
Ayele, E.e.a.: Internet of things for environmental monitoring: A review. Environmental Monitoring and Assessment  \textbf{192}(5),  1--21 (2020)

\bibitem{bainomugisha2023design}
Bainomugisha, E., Ssematimba, J., Okure, D.: Design considerations for a distributed low-cost air quality sensing system for urban environments in low-resource settings. Atmosphere  \textbf{14}(2), ~354 (2023)

\bibitem{bainomugisha2024ai}
Bainomugisha, E., Warigo, P.A., Daka, F.B., Nshimye, A., Birungi, M., Okure, D.: Ai-driven environmental sensor networks and digital platforms for urban air pollution monitoring and modelling. Societal Impacts  \textbf{3},  100044 (2024)

\bibitem{bigquery}
BigQuery, G.: Google bigquery. Available at \url{https://cloud.google.com/bigquery} (Accessed on August 19, 2025) (2025)

\bibitem{cai2015challenges}
Cai, L., Zhu, Y.: The challenges of data quality and data quality assessment in the big data era. Data science journal  \textbf{14}, ~2--2 (2015)

\bibitem{chhikara2021federated}
Chhikara, P., Tekchandani, R., Kumar, N., Tanwar, S., Rodrigues, J.J.: Federated learning for air quality index prediction using uav swarm networks. In: 2021 IEEE global communications conference (GLOBECOM). pp.~1--6. IEEE (2021)

\bibitem{dushyanth2025design}
Dushyanth, V., Chakravarthi, R., Chaudhary, P., Kandhari, H., Kuanr, M., Dev, S.: Design and implementation of a low-power wireless sensor network for environmental monitoring in iot environments. In: 2025 International Conference on Automation and Computation (AUTOCOM). pp. 679--684. IEEE (2025)

\bibitem{endres2022synthetic}
Endres, M., Mannarapotta~Venugopal, A., Tran, T.S.: Synthetic data generation: A comparative study. In: Proceedings of the 26th international database engineered applications symposium. pp. 94--102 (2022)

\bibitem{garg2013apache}
Garg, N.: Apache kafka. Packt Publishing Birmingham, UK (2013)

\bibitem{van2014trans}
Van~de Giesen, N., Hut, R., Selker, J.: The trans-african hydro-meteorological observatory (tahmo). Wiley Interdisciplinary Reviews: Water  \textbf{1}(4),  341--348 (2014)

\bibitem{haines2022workflow}
Haines, S.: Workflow orchestration with apache airflow. In: Modern Data Engineering with Apache Spark, pp. 255--295. Springer (2022)

\bibitem{harenslak2021data}
Harenslak, B.P., De~Ruiter, J.: Data pipelines with apache airflow. Simon and Schuster (2021)

\bibitem{hashmy2023modular}
Hashmy, Y., Khan, Z.U., Ilyas, F., Hafiz, R., Younis, U., Tauqeer, T.: Modular air quality calibration and forecasting method for low-cost sensor nodes. IEEE Sensors Journal  \textbf{23}(4),  4193--4203 (2023)

\bibitem{higashino2017edge}
Higashino, T., Yamaguchi, H., Hiromori, A., Uchiyama, A., Yasumoto, K.: Edge computing and iot based research for building safe smart cities resistant to disasters. In: 2017 IEEE 37th international conference on distributed computing systems (ICDCS). pp. 1729--1737. IEEE (2017)

\bibitem{kumar2015rise}
Kumar, P., Morawska, L., Martani, C., Biskos, G., Neophytou, M., Di~Sabatino, S., Bell, M., Norford, L., Britter, R.: The rise of low-cost sensing for managing air pollution in cities. Environment international  \textbf{75},  199--205 (2015)

\bibitem{openaq}
Lewis, C.H., et~al.: Openaq: Building a global community around open air quality data. Data Science Journal  \textbf{19},  1--13 (2020)

\bibitem{lipovac2024developing}
Lipovac, I., Babac, M.B.: Developing a data pipeline solution for big data processing. International Journal of Data Mining, Modelling and Management  \textbf{16}(1),  1--22 (2024)

\bibitem{luigi}
Luigi: Luigi: Python module that helps build complex pipelines of batch jobs. Available at \url{https://luigi.readthedocs.io/} (Accessed on August 19, 2025) (2025)

\bibitem{mberu2025urban}
Mberu, B.U., Nsoesie, E.O., Abbasi, M., Adrine, P., Aiyegbajeje, F.O., Asemadahun, E., Bainomugisha, E., Bakibinga, E., Bakibinga, P., Beguy, D., et~al.: Urban health in africa  (2025)

\bibitem{munappy2020data}
Munappy, A.R., Bosch, J., Olsson, H.H.: Data pipeline management in practice: Challenges and opportunities. In: International Conference on Product-Focused Software Process Improvement. pp. 168--184. Springer (2020)

\bibitem{ndembi2025integrating}
Ndembi, N., Rammer, B., Fokam, J., Dinodia, U., Tessema, S.K., Nsengimana, J.P., Mwangi, S., Adzogenu, E., Dongmo, L.T., Rees, B., et~al.: Integrating artificial intelligence into african health systems and emergency response: Need for an ethical framework and guidelines (2025)

\bibitem{okure2025case}
Okure, D., Bainomugisha, E., Ogenrwot, D., Sserunjogi, R., Adrine, P., Okello, G.: Case study of participatory data-driven approaches to improve urban air quality in kampala, uganda. Urban Health in Africa p.~255 (2025)

\bibitem{okure2022characterization}
Okure, D., Ssematimba, J., Sserunjogi, R., Gracia, N.L., Soppelsa, M.E., Bainomugisha, E.: Characterization of ambient air quality in selected urban areas in uganda using low-cost sensing and measurement technologies. Environmental Science \& Technology  \textbf{56}(6),  3324--3339 (2022)

\bibitem{openweathermap}
OpenWeatherMap: Weather model - openweathermap. Available at \url{https://openweathermap.org/technology/} (Accessed on August 19, 2025) (2012)

\bibitem{pinder2019opportunities}
Pinder, R.W., Klopp, J.M., Kleiman, G., Hagler, G.S., Awe, Y., Terry, S.: Opportunities and challenges for filling the air quality data gap in low-and middle-income countries. Atmospheric environment  \textbf{215},  116794 (2019)

\bibitem{prefect}
Prefect: Prefect: The modern data workflow orchestration. Available at \url{https://www.prefect.io/} (Accessed on August 19, 2025) (2025)

\bibitem{raj2020modelling}
Raj, A., Bosch, J., Olsson, H.H., Wang, T.J.: Modelling data pipelines. In: 2020 46th Euromicro conference on software engineering and advanced applications (SEAA). pp. 13--20. IEEE (2020)

\bibitem{redshift}
Redshift, A.: Amazon redshift - cloud data warehouse. Available at \url{https://aws.amazon.com/redshift/} (Accessed on August 19, 2025) (2025)

\bibitem{Singh2019}
Singh, P.: Learn PySpark: Build Python-based machine learning and deep learning models. Apress (2019)

\bibitem{iotpipelines}
Smith, J.e.a.: Scalable data pipelines for iot sensor streams. IEEE Internet of Things Journal  \textbf{8}(3),  2001--2015 (2021)

\bibitem{snowflake}
Snowflake: Snowflake. Available at \url{https://www.snowflake.com/} (Accessed on August 19, 2025) (2025)

\bibitem{sserunjogi2022seeing}
Sserunjogi, R., Ssematimba, J., Okure, D., Ogenrwot, D., Adong, P., Muyama, L., Nsimbe, N., Bbaale, M., Bainomugisha, E.: Seeing the air in detail: Hyperlocal air quality dataset collected from spatially distributed airqo network. Data in brief  \textbf{44},  108512 (2022)

\bibitem{thingspeak}
ThingSpeak: Iot analytics - thingspeak internet of things. Available at \url{https://thingspeak.com/} (Accessed on August 19, 2025) (2025)

\bibitem{ullah2024ai}
Ullah, U., Usama, M., Muhammad, Z., Akbar, A.: Ai-enabled low-powered wireless area networks for quality air. In: Low-Power Wide Area Network for Large Scale Internet of Things, pp. 100--141. CRC Press (2024)

\bibitem{wang2015building}
Wang, G., Koshy, J., Subramanian, S., Paramasivam, K., Zadeh, M., Narkhede, N., Rao, J., Kreps, J., Stein, J.: Building a replicated logging system with apache kafka. Proceedings of the VLDB Endowment  \textbf{8}(12),  1654--1655 (2015)

\bibitem{zafra2024designing}
Zafra-P{\'e}rez, A., Medina-Garc{\'\i}a, J., Boente, C., G{\'o}mez-Gal{\'a}n, J.A., de~la Campa, A.S., De~La~Rosa, J.: Designing a low-cost wireless sensor network for particulate matter monitoring: Implementation, calibration, and field-test. Atmospheric Pollution Research  \textbf{15}(9),  102208 (2024)

\bibitem{zhalgasbekova2017opportunistic}
Zhalgasbekova, A., Zaslavsky, A., Saguna, S., Mitra, K., Jayaraman, P.P.: Opportunistic data collection for iot-based indoor air quality monitoring. In: International Conference on Next Generation Wired/Wireless Networking. pp. 53--65. Springer (2017)

\end{thebibliography}
\end{document}